\documentclass[prl,aps,amsmath,amssymb,amsfonts,twocolumn,superscriptaddress,nofootinbib,amssymb]{revtex4-1}

\UseRawInputEncoding

\usepackage{xcolor}
\usepackage{amsmath}
\usepackage{bbm}	
\usepackage{mathtools}
\usepackage{siunitx}
\usepackage{tikz}
\usepackage{times}
\usetikzlibrary{calc,matrix,positioning,fit,patterns,decorations.text,fadings}

\definecolor{jen}{rgb}{0,0,0}
\newcommand{\je}[1]{{\color{jen} #1}}

\newcommand{\OL}[1]{{\color{jen} #1}}
\definecolor{newjens}{rgb}{0,0,0}

\newcommand{\acom}[2]{\{#1,#2\}}

\newcommand{\ket}[1]{|#1\rangle}

\newcommand{\onerdm}{\rho^{(1)}}

\begin{document}
\title{\je{Effective} dimension reduction with mode transformations: \\
Simulating
two-dimensional fermionic condensed matter systems
with matrix-product states}
\author{C.\ Krumnow}
\affiliation{Dahlem Center for Complex Quantum Systems, Freie Universit{\"a}t Berlin, 14195 Berlin, Germany}
\author{L.\ Veis}
\affiliation{J. Heyrovsk\'y Institute of Physical Chemistry, Academy of Sciences of
the Czech Republic, Dolej\v{s}kova 3, 18223 Prague 8, Czech Republic}
\author{{J.\ Eisert}}
\affiliation{Dahlem Center for Complex Quantum Systems, Freie Universit{\"a}t Berlin, 14195 Berlin, Germany}
\affiliation{Department of Mathematics and Computer Science, Freie Universit{\"a}t Berlin, Germany}
\author{{\"O}.\ Legeza}
\affiliation{Wigner Research Centre for Physics, Hungarian Academy of Sciences, Budapest, Hungary}

\begin{abstract}
Tensor network methods have progressed from variational techniques based on matrix-product states able to compute properties of one-dimensional condensed-matter lattice models into methods rooted in more elaborate states such as projected entangled pair states aimed at simulating the physics of two-dimensional models. In this work, we advocate the paradigm that for two-dimensional fermionic models, matrix-product states are still applicable to significantly higher accuracy levels than direct embeddings into one-dimensional systems allow for. To do so, we exploit schemes of fermionic mode transformations and overcome the prejudice that one-dimensional embeddings need to be local. This approach takes the insight seriously that the suitable exploitation of both the manifold of matrix-product states and the unitary manifold of mode transformations can more accurately capture the natural correlation structure. By demonstrating the residual low levels of entanglement in emerging modes, we show that matrix-product states can describe ground states strikingly well. The power of the approach is exemplified by investigating a phase transition of spin-less fermions for lattice sizes up to $10\times 10$.
\end{abstract}
\maketitle

Recent years have enjoyed a flourishing development of tensor network methods, entanglement-based methods that allow
to describe strongly correlated quantum many-body systems \cite{OrusReview,VerstraeteBig,EisertReview,Szalay-2015,AreaLaw}. 
They originate from the powerful density-matrix renormalization group (DMRG) 
\cite{White-1992b,White-1992a,Schollwoeck2011},
a variational method building on matrix-product states (MPS) \cite{Fannes-1992,Perez-Garcia-2007,Rommer-1997}
that captures the physics of one-dimensional local Hamiltonian 
systems provably well \cite{Hastings-2007,1301.1162,VerstraeteBig,AreaSchuch}.  
It has been applied to countless physical systems (see the reviews \cite{Schollwock-2005,Schollwoeck2011} and the comprehensive 
web page \cite{NishinoWeb}) and extended to time-evolving systems \cite{PhysRevLett.91.147902,Daley2004,Haegeman-2014b}, 
open systems \cite{Mixed,PositiveMPO}, and
the study of excited states \cite{PhysRevLett.116.247204}. 
Generalizing the variational set of matrix-product states to projected entangled pair states in two spatial dimensions,
new avenues for the study of strongly correlated systems with tensor networks followed \cite{PEPSOld,VerstraeteBig,OrusReview}, including
studies of fermionic models 
\cite{PhysRevA.80.042333,SchuchFermiPEPS,PhysRevB.81.245110,PhysRevB.93.045116}.

\je{Interestingly}, even if the DMRG approach has originally been devised to capture one-dimensional systems only: There are regimes in which it interestingly still performs competitively well \cite{KagomeHeisenberg,PhysRevLett.109.067201}
even in situations that at first seem alien to that type of approach and in which area laws for entanglement entropies
are violated \cite{AreaLaw}. Two-dimensional strongly correlated systems can be
naturally embedded in highly non-local Hamiltonian models on a line. The high degree of entanglement that renders a variational approach based on
matrix-product states challenging are partially compensated by the facts that contraction is efficient, and that very large bond dimensions
are accessible. DMRG produces relevant data for strongly correlated matter even in two spatial dimensions, and for systems with
fermionic degrees of freedom \cite{Stoudenmire-2012}.
The significance of this insight is even strengthened by the fact that 
DMRG is strictly variational, so that all ground state 
energies generated are \je{precisely upper} 
bounds. And yet, given that the
entanglement structure is not fully captured by matrix-product states,
there are strong limitations of direct DMRG approaches.

In this work, we bring the idea of 
tackling two-dimensional strongly correlated matter with one-dimensional matrix-product states 
to a new level.
We show that the potential of 
\je{using one-dimensional tensor network
states for classically simulating
higher
dimensional quantum systems} \je{-- 
in what we refer to as an 
effective dimension reduction in the description 
--}
is significantly more powerful than anticipated. We do so by systematically exploiting a degree of freedom that has not sufficiently been appreciated in the study of strongly correlated condensed-matter systems: This is the degree of freedom to adaptively define suitable modes in a strongly correlated fermionic system. Its significance is already manifest when solving problems in either real or in momentum space 
\cite{Xiang-1996,Nishimoto-2012,Legeza-2003b,Legeza-2006b,Murg-2010a,Ehlers-2015,Motruk-2016,Ehlers-2017}. 
For $n$ fermionic
modes, however, there is an entire $U(n)$ freedom that can be made use of and exploited when devising variational
principles. In fact, a manifold structure emerges that originates from the tensor network and mode transformation degrees of
freedom. Only the joint optimization fully exploits the potential of matrix-product state approaches in the study of
strongly correlated fermionic condensed-matter system.
It is this serious gap in the literature that is closed in this work: We overcome the prejudice that a one-dimensional embedding 
necessarily has to be an embedding in 
real space. \je{We come to this conclusion not only based at hand of the evidence of substantially improved energies. We also find that the mode-optimized quantum states indeed feature one-dimensional
entanglement area laws.}

\emph{Setting.}
The Hilbert space of interacting fermions in $n$ modes
is the fermionic Fock space $\mathcal{F}_n$
originating from the basis constituted by all Slater determinants $\{\ket{\alpha_1,\dots,\alpha_n}\}$ with $\alpha_j\in\{0,1\}$.
We denote with $c_j$ the fermionic annihilation operator of mode $j$ satisfying the canonical anti-commutation relations $\acom{c_i}{c_j}=0$ and $\acom{c_i^\dag}{c_j}=\delta_{i,j}$.
MPS vectors then take the form
	\begin{equation} \ket{\psi} = \sum\limits_{\alpha_1,\ldots,\alpha_n=1}^d A^{\alpha_1}_{[1]}\ldots A^{\alpha_n}_{[n]}\ket{\alpha_1\dots\alpha_n}.\label{eq:DefinitionMPS}
	\end{equation}
We build upon ideas of adaptive fermionic mode transformations \cite{PhysRevLett.117.210402,phd,Krumnow-2019}, here brought to the
level of applicability to condensed-matter lattice models in two spatial
dimensions. 
To be specific, and to exemplify the power of our approach, the example of the spin-less interacting fermionic (spin-less Fermi-Hubbard)
model 
\begin{equation}
 H = \sum\limits_{\langle i,j\rangle} c_i^\dag c_j + 
        \sum\limits_{\langle i,j\rangle} V n_i n_j,
\end{equation}
will be in the focus of attention, where $V$ is the interaction strength, the hopping amplitude is set to 1, $n_j= c_j^\dagger c_j$, 
and $\langle i,j\rangle$ denotes nearest neighbours $i,j\in[n]$
on a \je{two-dimensional}  cubic $N\times N$ lattice with $n=N^2$. Periodic boundary condition will be 
imposed along both spatial dimensions, which has been considered as a major
bottleneck for MPS-based approaches.
This example will show-cast that state-of-the-art energies can be reached.
Having said that, in the mindset of this work would be
any translationally invariant Hamiltonian of the form 
\begin{equation} H = \sum\limits_{i,j=1}^{n} t_{i,j} c_i^\dag c_j + \sum\limits_{i,j,k,l=1}^{n} v_{i,j,k,l}c_i^\dag c_j^\dag c_l c_k, \label{eq:Hamiltonian}\end{equation} 
including local spin degrees of freedom. That is to say, the Hamiltonian is
treated as a long-ranged fermionic model on a one-dimensional line equipped with a given ordering. 

\emph{Methods.} We optimize the single particle basis in conjunction with the MPS tensors withing multiple successive mode transformation iterations. \je{We refer this prodecure leading to a 
state-of-the-art variational
ground state approximation with one-dimensional
tensor networks as an
effective dimension reduction
in the description of a higher-dimensional fermionic
system.}
In our implementation, a single mode transformation iteration consists of 
a full forward and backward DMRG sweep without basis rotations 
using the \emph{dynamically extended active 
space (DEAS)} procedure \cite{Legeza-2003b,Szalay-2015}, which is
followed by some number of additional sweeps with local mode transformations that adapt the single particle basis (compare Refs.~\cite{PhysRevLett.117.210402,phd}) that also rotate the couplings in the Hamiltonian to general couplings $t_{i,j}^\prime$ and $v_{i,j,k,l}^\prime$.
At the end of the 
last sweep, for the symmetric
super-block configuration, we have calculated the site entropies $s_i$, the 
two-site \emph{mutual information} $I_{i,j}=s_i+s_j-s_{i,j}$,
the \emph{one-particle reduced density matrix}, $\onerdm$, and the \emph{occupation
number distribution} $\langle n_i\rangle$ with $i\in\{1,\ldots , n\}$.
Here $s_{\rm A}= -{\rm Tr} (\rho_{\rm A} \ln \rho_{\rm A})$ for $A\subset[n]$
is the von-Neumann entropy of the reduced state
obtained from a partial trace of the full 
quantum state. 
The eigenvalues of $\onerdm_{i,j} = \langle c^\dagger_i c_j\rangle$ define the \emph{natural occupation (NO)} numbers, 
$\lambda_i$, and its eigenvectors
the NO-basis. 
Based on $I_{i,j}$ we have calculated an optimized ordering
using the Fiedler-vector approach \cite{OrsQI}, from $\{s_i\}$ a new
{complete active space vector} for the DEAS procedure \cite{Legeza-2003b} and from $\langle n_i\rangle$ a new Hartree-Fock configuration. 
These together with the final rotated
interaction matrices 
are all used  
as inputs for the subsequent mode transformation iteration. 

The basis optimization \je{has been} carried out with fixed low
bond dimension $D_{\rm opt}\simeq 64$ and $256$ or with a 
systematic increase of $D_{\rm opt}$ as will be discussed below.
After convergence is reached 
large scale DMRG calculations are performed with increasing
bond dimension or using the \emph{dynamic block state selection
(DBSS)} approach with fixed truncation error threshold~\cite{Legeza-2003a,Legeza-2004b}. 
We denote these data as $(D_{\rm opt},D)$ or $(D_{\rm opt},\varepsilon_{\rm tr})$, respectively.  %
In addition, a given quantity obtained 
from a calculation in the optimized basis will be indicated with a
tilde.
\OL{In the supplements, further results with $D_{\rm opt}$ up to 1024 are also discussed.}

\emph{Numerical results.} 
Our systematic error and convergence analysis will be given for the
$6\times6$ two dimensional lattice, since highly accurate
reference data with the real space basis can also be generated.
For larger system sizes, namely for $8\times8$ and $10\times10$, only
final results will be discussed
(further numerical \OL{aspects,} data and figures are presented
in the supplements).

\begin{figure}[t]
\includegraphics[width=0.98\columnwidth]{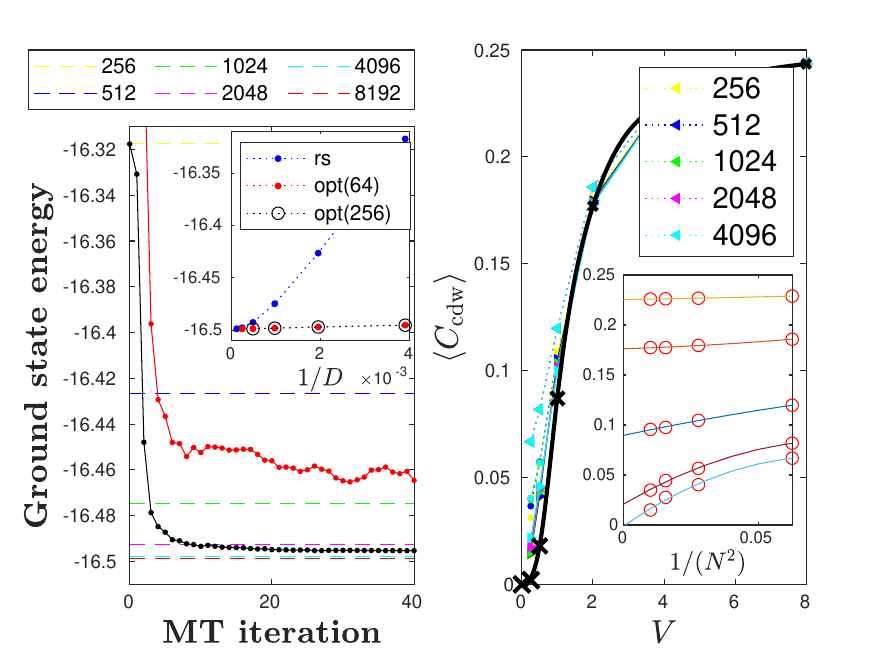}
\caption{(Left panel) convergence of the ground state energy 
for the half-filled $6\times6$ spin-less fermion model for $V=1$ 
as a function of
mode transformation iterations for fixed bond dimension of 
$D_{\rm opt}=64$ and $256$ is shown by red and black curves, respectively. 
In the inset, the scaling of the ground state energy with inverse 
bond dimension obtained in the real space basis 
and in the optimized basis for $D_{\rm opt}=64$ and $256$ are shown, respectively.
(Right panel) Charge density wave order parameter 
for the $N\times N$ half-filled spin-less fermion model
as a function of $V$
for various values of $D$ obtained in the real space basis.
Black crosses indicate extrapolated data to the $N\rightarrow\infty$
limit obtained in the optimized basis and using 
finite size scaling data shown for various interaction strengths 
in the inset {(curves correspond from bottom to top to $V=0.25$, 
$0.5$, $1$, $2$, $4$, and $8$)}. The 
solid black line is a spline fit to the extrapolated data. 
The error in the extrapolated data is indicated by the symbol sizes.}
\label{fig:energy-cdw-6x6}
\end{figure}

In the left panel of Fig.~\ref{fig:energy-cdw-6x6}, we show the 
ground state energy $\tilde E(D_{\rm opt})$ for $V=1$ as a 
function of mode transformation iterations using fixed bond dimensions 
$D_{\rm opt}=64$ and $256$. Reference energies $E(D)$ obtained in the real 
space basis are indicated with dashed lines for various bond dimensions
up to $D=8192$. It obvious indeed 
that exploiting mode transformations, $\tilde E(64)$ gets
significantly below $E(512)$ even after the fourth 
iteration step and $\tilde E(256)$ is below $E(2048$). 
For further numerical results emphasizing how faithfully information beyond the ground state energy can be reproduced and predicted in the optimized basis,
we refer to Fig.~\ref{fig:rho-6x6} in the supplements.
In the inset of the left panel of Fig.~\ref{fig:energy-cdw-6x6}, we depict the ground state energy
as an inverse of the bond dimension for the real space basis and for
the optimized basis with $D_{\rm opt}=64$ and $256$. In the latter case,  
$\tilde E(D_{\rm opt},D)$ lie on the top of each other, 
indicating that the optimal basis has been 
found with $D_{\rm opt}=64$ already 
(red dots in black circles).

For larger system sizes, the improvements are even more remarkable
as is shown in Fig.~\ref{fig:energy-8x8} in the supplements
for the $8\times 8$ lattice for different values of  
$D_{\rm opt}$ and for $V=1$ and $8$.  
Here, $\tilde E(256)$ is already lower than $E(8192)$.   
In addition, reliable extrapolation with $1/D$
to the $D\rightarrow\infty$ truncation free limit would require 
even significantly larger bond dimensions for the real space basis. 
In contrast to this, in case of 
the optimized basis, this is no longer an issue since  
$\tilde E(256, D)$ is basically a flat curve.
Our very accurate results have been obtained for 
a torus geometry. This reduces finite size effects significantly 
and much smaller systems sizes could lead to a reliable extrapolation
to the thermodynamic limit (see 
Tab.~\ref{tab:bondenergy}).

\begin{figure}[t]
\includegraphics[width=1.0\columnwidth]{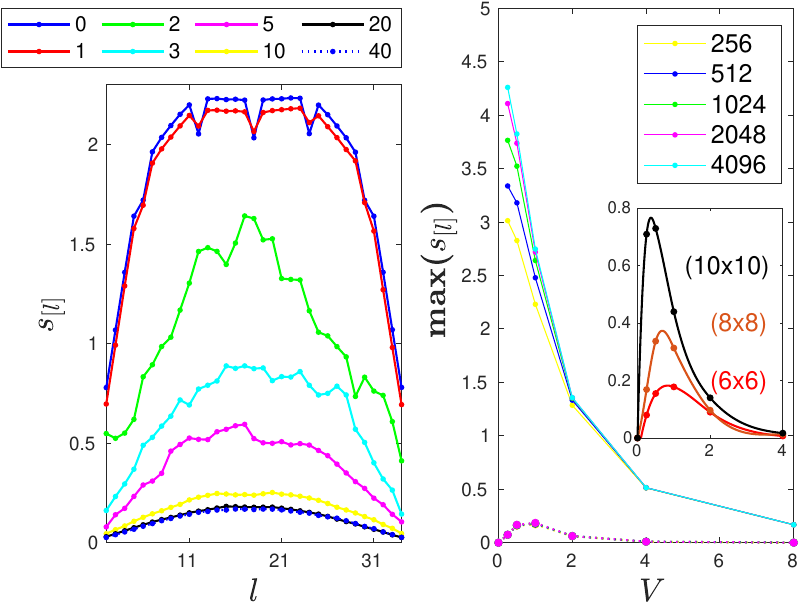}
\caption{
(Left panel) block entropy for
the $6\times6$ half-filled spin-less fermion model for $V=1$
for some selected mode transformation iterations with $D_{\rm opt}=256$, 
i.e., 
for the 0$^{\rm th}$, 1$^{\rm th}$, 2$^{\rm nd}$, 3$^{\rm rd}$, 5$^{\rm th}$, 10$^{\rm th}$, 20$^{\rm th}$, 40$^{\rm th}$ iterations.
(Right panel) maximum of the block entropy 
as a function of $V$ for various $D$ values and for the real space basis (solid lines) and for the optimized basis (dashed lines). In the inset, this is shown for various systems sizes
obtained with the optimized basis and DMRG with $D_{\rm min}=1024$ and $\varepsilon_{\rm tr}=10^{-6}$. Here a spline is fitted as guide for the eye trough the data points.
}
\label{fig:entropy-6x6}
\end{figure}

The remarkable superiority of the optimized basis over the real space
basis is due to the dramatic reduction of the entanglement.
As an indication of this, we depict the block
entropy $s_{[l]}$, $l\in\{1,\ldots,n\}$ in the left panel of
Fig.~\ref{fig:entropy-6x6} for
various selected mode transformation iterations.
Here, the maximum of $s_{[l]}$ reduced by a full 
order of magnitude, as can be seen
by comparing the blue (real space basis) and the black (optimized basis)
curves. In addition, artifacts of the snake-like mapping of the two-dimensional
lattice in real space
into the one-dimensional MPS topology apparent in the blue curve
are completely diminished by the basis optimization resulting
in a smooth and highly symmetric profile (additional data is available in the 
\je{supplements)}.
The iterative error norm of the block entropy measured between two subsequent
mode transformation iterations,
$\|s_{[l]}^{k+1} - s_{[l]}^{k}\|$ converges 
to $10^{-5}$-10$^{-4}$ which can also be used as a criterion when 
to terminate the basis optimization.   
For larger $V$ values, the reduction is even more pronounced, 
leading to 
a state that is close to a Slater determinant. In the right panel, 
the maximum of 
$s_{[l]}$ for $l\in\{1,\dots,n\}$ -- which typically appears near the center of the chain -- is laid out
for various $D$ values
for the real space basis and for the optimized one. While
a strong $D$ dependence for $V\leq 2$ is clearly visible 
in the real space basis, the curves basically 
fall on top of each other for the optimized basis.  
The small peak for $0\leq V \leq 2$ signals the residual entanglement
that cannot be removed by basis optimization 
which also controls the required bond dimension and thus the 
computational complexity.
As a benchmark we have performed DMRG calculations using the
DBSS approach 
with minimum bond dimension 
$D_{\rm min}=1024$ 
and a truncation error
threshold $\varepsilon_{\rm tr}=10^{-7}$. 
An agreement up to four digits has been obtained
compared to the real space energy reference data calculated with 
$D=8192$, but we have gained a speedup by a full 
order of magnitude. 
\je{Strikingly, as we discuss in great detail in the 
supplements, while there are complexity
theoretic obstructions against a mapping of the given Hamiltonian to a local one-dimensional gapped Hamiltonian, the mode-transformed states are numerically found to feature an 
entanglement entropy that is upper bounded by a constant in the system size and the bond dimension, providing further strong evidence for the significance of the effective dimension reduction in description}
\OL{as shown in Figs.~\ref{fig:6x6Entropy} and \ref{fig:8x8Entropy}.} 

\emph{Phase diagram.}
The power of our approach allows us to attack the physical properties such as the phase diagram of the system as well.
In the limit of strong interactions, 
the model maps onto the anti-ferromagnetic Ising model in two dimensions and a \emph{charge density wave (CDW)} 
phase develops. 
Since the hopping is restricted to nearest neighbours only, the Fermi surface takes the form of a square and perfect nesting together with Van Hove singularities providing strong arguments for an Ising transition into the CDW-ordered phase at $V_{\rm c}=0$ 
\cite{solyom-2011,Metzner-2012}.
Furthermore, investigations within the Hartree-Fock approximation
lead to an exponentially small 
order parameter in the weak coupling limit and the $1/d$ corrections starting from the $d=\infty$ limit, where Hartree-Fock 
theory becomes exact, provides only very small quantitative corrections in $d=3$ and even in $d=2$ \cite{Halvorsen-1994}.
For $d=2$ this indicates a transition at $V_{\rm c}=0$ and that the charge density wave order parameter is an 
exponential function of $V$ in the weak coupling limit.
Note, however, that these simple arguments can break down as in the case of spin-less fermions in one 
spatial dimension, $d=1$,
where the model 
reduces to the integrable Heisenberg model and has a transition at finite $V_{\rm c}$ \cite{Halvorsen-1994}.  
Ref.~\cite{Uhrig-1993} has shown that  
there is a direct transition between the homogeneous 
and the CDW phases governed by phase separation, and a finite $V_c\simeq 0.5$
is suggested based on their obtained phase diagram.
Their underlying arguments, however, have been derived for finite 
doping, thus an exponentially closing phase boundary between the CDW 
and phase separated
phases together with $V_{\rm c}=0$ cannot be ruled out.  

In order to investigate the transition, we first analyze the block entropy profiles
for larger system sizes
using the optimized basis and find that the peak for $V\leq 1$  
remains and its height increases
with system size as is shown in \OL{the inset of the right panel of} Fig.~\ref{fig:entropy-6x6}. 
The center of the peak extracted from the spline fits ($V=0.83, 0.65, \OL{0.36}$ for $N=6, 8, 10$)
tends to shift to $V=0$ with $1/N^2$
\OL{which indicates a quantum phase transition} \cite{Legeza-2006a} at
$V_{\rm c}=0$. 
We also compute the CDW order parameter \cite{Wang-2014} as expectation value of
$C_{\rm cdw} = (1/N^4)\sum_{i,j} \eta_{i,j} (n_i-1/2)(n_j-1/2)$ directly, 
where  $n_i=c^\dag_i c_i$ in the real space basis
and $\eta_{i,j}$ is a phase matrix with elements $\pm 1$ in a checker-board arrangement on the \je{two-dimensional} lattice. 
The real-space simulations show that for large values of $V$, $\langle C_{\rm cdw}\rangle$ takes a finite value while for $V=0$ it has to vanish as can be seen in the right panel of Fig.~\ref{fig:energy-cdw-6x6}.
The apparent finite size and $D$ dependencies do not allow
us to conclusively decide upon the behaviour of $\langle C_{\rm cdw}\rangle$ for $V\leq 1$.
Alternatively, the density-density
correlation function can also be taken from the elements of the
one- and two-particle reduced density matrices. The latter one
has entries $\rho^{(2)}_{i,j,k,l}=\langle c^\dagger_i c^\dagger_j c_k c_l\rangle$ which can also be calculated efficiently by the DMRG method~\cite{Zgid-2008b}.
Measuring these in the optimized basis and back-rotating 
to the real space basis,
we \je{have} found an agreement up to four digits between 
$\langle \tilde C_{\rm cdw}\rangle$ and the real space reference for 
$N=4$ and $6$. For $N=8$ and $V\leq 1$ the two data sets, however, 
began to deviate and 
$\langle\tilde C_{\rm cdw}\rangle$ possesses a much weaker $D$ dependence.
Finite size scaling of the large scale DMRG data obtained with 
$M_{\rm min}=1024$ and $\varepsilon_{\rm tr}=10^{-6}$ is shown in 
the inset of Fig.~\ref{fig:energy-cdw-6x6} right panel for various $V$ values.
For large $V$ the curves scale to finite values in the thermodynamic limit,
while for $V\leq 1$ they show a slight downward curvature. 
After a rough extrapolation with $1/N$ 
and a spline fit on the extrapolated data (black crosses in the figure) an exponential 
opening of $\langle \tilde C_{\rm cdw}\rangle$ at $V_{\rm c}=0$ 
\OL{has been obtained}.
This functional form agrees to the one reported in 
Ref.~\cite{Halvorsen-1994} after some
re-scaling and it is shown by a black curve in the right panel of Fig.~\ref{fig:energy-cdw-6x6}. 
Our approach hence pushes forward the
capacity of the MPS based approaches to capture two dimensional 
strongly correlated systems significantly.
Our results are in close agreement with analytic expectations (while some details remain open).

\emph{Conclusion.} In this work, we have demonstrated that 
MPS approaches, extending known  DMRG methods, are
surprisingly powerful for
the simulation of two-dimensional 
quantum many body systems even imposing periodic boundary
condition along both spatial dimensions. This is possible if only
the key insight is acknowledged that one is not forced to do a local
basis representation. Algorithmically,
this is achieved by adaptively finding the 
optimal basis via fermionic mode transformation, optimizing over
a larger manifold than that of MPS, which leads to a dramatic
reduction of the correlations and entanglement in the system.
A strongly interacting model in the real space basis thus
can be converted to a weakly correlated problem in the
optimized basis. Due to the torus geometry,
finite size dependence is significantly reduced and intermediate system
sizes make it possible to carry out more reliably extrapolations
to the thermodynamic limit.
{In fact, for the two-dimensional translationally invariant 
spin-less fermion model,
our results strongly suggest the presence of a 
quantum phase transition at $V_{\rm c}\simeq 0$, but
the very small values of the charge density order parameter obtained numerically 
in the weak coupling limit leaves an uncertainty in our conclusion. 
The inclusion of a hopping between next nearest neighbours, however, would distort the square
Fermi-surface and 
perfect nesting over an extended region of the momentum space will be destroyed.
This is expected to have a have major effect,
and divergencies in the susceptibilities might be removed
and a finite $V_{\rm c}$ is even more likely. 
This behaviour also shares features with the phase diagram of 
spin-less fermions on the honeycomb lattice \cite{Sylvain-2017}.
Then, physical properties of the transformed basis
are of key importance.
In general, the ground state energy cannot be written as a sum 
of energies of quasi-particle states except for special cases. 
The $V=0$ and large $V$ limits belong to the latter case 
(the ground state  is a product state),
but the residual block entropy for $0<V\leq 2$ reflects \je{the general
scenario}.} 

Our basis optimization is very robust, it can be carried out with 
low bond dimension, and calculations using the optimized basis
can easily lead to an order of magnitude speedup in computational time.
In addition, our method is stable for weakly and strongly interacting
systems, in general, while standard approaches, like basis transformation
based on natural orbitals, that have been attempted 
earlier \OL{\cite{Rissler-2006}} have major limitations 
and drawbacks
(for numerical data see Fig.~\ref{fig:mt-iter-8x8_NOenergy}).
Remarkably, the optimized basis for the spin-full Hubbard model does not resemble the characteristics of 
natural orbitals which reflects the existence of much stronger residual correlations in the system (as forthcoming work will explore).
Conceptually most importantly, our work
overcomes the deep misconception that lower-dimensional
embeddings necessarily have to capture some kind of locality.
Once this prejudice is overcome, acknowledging that 
fermionic mode transformations are not restricted to one-dimensional
embeddings, mode 
\je{transformations} and \je{effective
dimension reductions in description}
can be brought to a new level.  
\je{This is even more interesting and surprising given that a full mapping on the level of Hamiltonians and their accompanying ground states to polynomial accuracy is in general not possible (as we 
elaborate on in more detail in the supplementary material).}
Due to the polynomial scaling of 
the non-local DMRG~\cite{white1999} effort as
$O(D^3n^3)+O(D^2n^4)$, a 
reduction of $D$ by one or two orders
of magnitude will render DMRG competitive 
for simulating
higher dimensional and complex problems as well.   
Our approach has the potential to become
a standard protocol for tensor network 
methods.

\emph{Acknowledgements.} J.~E.\ has been supported by the 
Templeton Foundation, the ERC (TAQ), 
the DFG (EI  519/15-1, EI 519/14-1, CRC 183 Project B01), 
and the European Union's Horizon 2020
research and innovation program under Grant Agreement  No.\  817482  (PASQuanS). 
\"O.~L.\ has  been  supported by the Hungarian National Research, Development and Innovation Office
(NKFIH) through Grant No.~K120569, by the Hungarian Quantum 
Technology National Excellence
Program (Project No. 2017-1.2.1-NKP-2017-00001). \"O.~L.~also 
acknowledges financial support
from the Alexander von Humboldt foundation and useful discussions 
with Jen{\H o} S\'olyom, Karlo Penc, and Florian Gebhard. 
L.~V.~has been supported by the Czech Science Foundation 
through Grant No.~18-18940Y.
\OL{The development of DMRG libraries has been supported by the Center for Scalable  and  Predictive  methods  for  Excitation  and  Correlated  phenomena  (SPEC),  which  is  funded  as  part  of  the Computational Chemical Sciences Program by the U.S. Department of Energy (DOE), Office of Science, Office of Basic  Energy  Sciences,  Division  of  Chemical  Sciences,  Geosciences, and Biosciences at Pacific Northwest National Laboratory.}
%

\bigskip \bigskip 
\appendix

\section{Supplemental material: Additional data for larger systems}

In this supplementary material, we present 
additional numerical data complementing the findings of the main text
together 
with further scaling properties obtained for 
larger system sizes \je{as well as further conceptual
considerations.}

\subsection{Error analysis of the one particle reduced density matrix}

\begin{figure}[!htb]
\includegraphics[width=1.0\columnwidth]{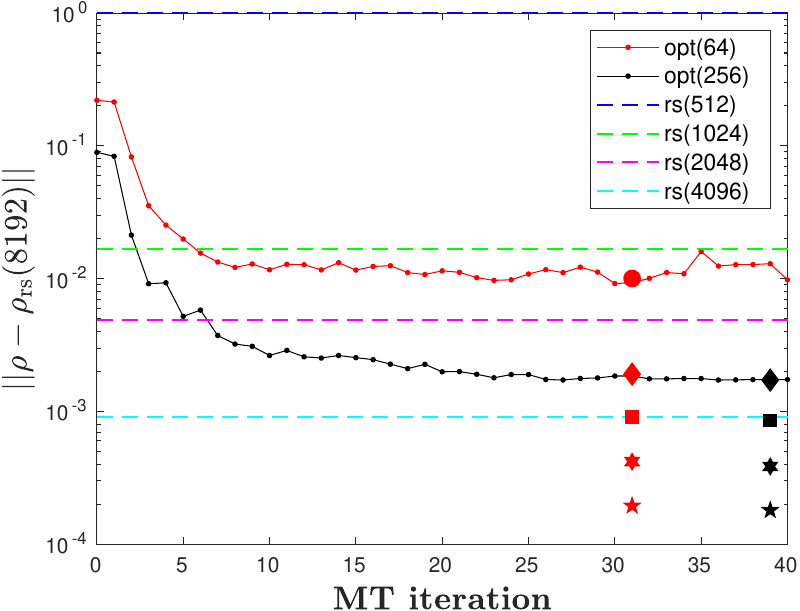}
\caption{
Error norm of the one-particle reduced density matrix 
with respect to the
reference obtained with the real space basis with $D=8192$. 
Red and black symbols show
result for $D=64,256,512, 1024, 2048$ but using the optimized basis for
$D_{\rm opt}=64$ and $256$, respectively.
For both quantities reference data
obtained with the real space basis for 
various $D$ values up to 8192 are shown with dashed lines 
and labeled as rs$(D)$. 
}
\label{fig:rho-6x6}
\end{figure}

In order to investigate how faithfully information beyond the ground state energy can be reproduced and predicted in the optimized basis, 
we depict in 
Fig.~\ref{fig:rho-6x6} 
the operator norm
of the difference of the one particle reduced 
density matrix $\widetilde{\onerdm}(D_{\rm opt})$ over the mode transformation iterations and the real space reference data $\onerdm(8192)$. 
Using the optimized basis, 
we also show the result 
for $\widetilde{\onerdm}(D_{\rm opt},D)$ with increasing bond dimension $D$, 
using different symbols. 
These latter data sets are basically the same for  
$D_{\rm opt}=64$ and $256$, thus the optimal basis has already been
obtained with the lower $D_{\rm opt}$ value (see Fig.~\ref{fig:rho-6x6}).
The error norms obtained with the real space basis are again much
larger as indicated by the dashed lines. 
The error norm is less meaningful for very large 
bond dimensions since $\tilde E(256, 4096)$ 
is below $E(8192)$ rendering $\widetilde{\onerdm}(256,4096)$ potentially more accurate than $\onerdm(8192)$.

\subsection{Further numerical results for the ground state energy of the half-filled $N\times N$ 
spin-less fermion model}

\begin{figure}[!htb]
\includegraphics[width=1.0\columnwidth]{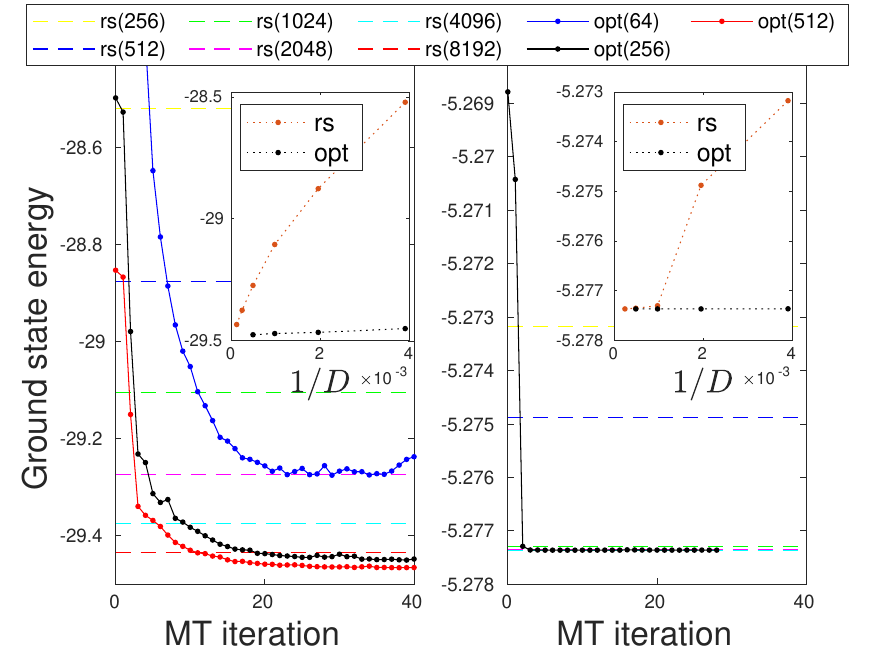}
\caption{Convergence of the ground state energy 
for the half-filled $8\times8$ spin-less fermion model  
as a function of
mode transformation iterations for fixed bond dimension of 
$D_{\rm opt}=64$, $256$ and $512$ is shown by blue, red and black curves, 
respectively for $V=1$ (left panel) and for $V=8$ (right panel). 
Reference data obtained in the real space basis for $D$ up to $8192$
are shown with dashed lines.
In the inset, the scaling of $E(D)$ and $\tilde{E}(256,D)$ with the inverse bond dimension is shown.
}
\label{fig:energy-8x8}
\end{figure}

{In Fig.~\ref{fig:energy-8x8}, we present further numerical results for the
ground state energy of the half-filled $8\times 8$ spin-less fermion model, and obtained bond energies are
summarized up to lattice sizes $10\times 10$ in Tab.~\ref{tab:bondenergy}.} 

\begin{table}[htb]
\begin{tabular}{|l|c|c|c|c|c|c|}
\hline
    &  0.25  &  0.5  &  1 & 2 & 4 & 8   \\
\hline
\hline
$4\times4$ &   -0.6636 &  -0.5845 &  -0.4518  & -0.2878 &  -0.1591  & -0.0823 \\
$6\times6$ &   -0.6847 &  -0.5992 &  -0.4583  & -0.2911 &  -0.1601  & -0.0825 \\
$8\times8$ &   -0.6947 &  -0.6059 &  -0.4606  & -0.2912 &  -0.1601  & -0.0825 \\
$10\times10$&  -0.69\OL{94} &  -0.60\OL{86} &  -0.46\OL{07}  & -0.2912 &  -0.1601  & -0.0825 \\
\hline
\end{tabular}
\caption{Convergence of the bond energy, $E/N^2$, with system size for various $V$ values. 
DMRG results were obtained using the optimized basis and the DBSS procedure with
$M_{\rm min}=1024$, and $\varepsilon_{\rm tr}=10^{-6}$.}
\label{tab:bondenergy}
\end{table}

\subsection{Complexity theoretic insights into dimensional reduction in description}

\je{The main point of this work is to provide evidence for the observation that an effective 
dimension reduction in description can lead to a
substantially improved classical simulation
of strongly correlated quantum systems: A one-dimensional tensor network ansatz can 
capture two-dimensional strongly correlated models well, if the prejudice is overcome
that the mapping to one spatial dimension has to be spatially local. The main text shows that such an effective dimension reduction on the level of description is possible,
leading to substantially improved descriptions over standard one-dimensional
embeddings. In the subsequent subsections, we provide further evidence for this at hand of discussing entanglement entropies and energies.

Having said that, we insist on an effective dimension reduction in description,
that is, on the level of the variational ansatz capturing the strongly correlated
quantum system at hand. On the level of Hamiltonians and concomitant ground states to polynomial accuracy, a full
efficient reduction to a one-dimensional
system is in general infeasible, as 
obstructions of computational complexity 
are in the way. In the light of this observation,
it is even more interesting that the 
effective dimension reduction in description
provides so convincing results.
%
%
In the following, we elaborate more on this conceptual obstruction.

Consider as input to the problem a family of
general strongly correlated fermionic systems
with Hamiltonian 
\begin{equation} H = \sum\limits_{i,j=1}^{n} t_{i,j} c_i^\dag c_j + \sum\limits_{i,j,k,l=1}^{n} v_{i,j,k,l}c_i^\dag c_j^\dag c_l c_k \label{eq:Hamiltonian_newdef}
\end{equation} 
as above. Under a mode transformation
defined by a $U\in U(n)$ such a
Hamiltonian transforms to a new local
(quartic) Hamiltonian of the same form,
albeit no longer necessarily a geometrically
local one even if the original Hamiltonian
has been geometrically local.
It has been shown in Ref.\ 
\cite{QMA} that to approximate the ground state
energy of such a model to polynomial
accuracy is {\tt QMA}-hard,
in a complexity-theoretic
language. This implies in particular that
it is ${\tt NP}$-hard, so no polynomial
time algorithm exists unless ${\tt NP}={\tt P}$. If
one could find a mode transformation
$U\in U(n)$ in  polynomial time 
that transforms the ground state to
a quantum state that is approximated by a 
matrix-product-state up to an error in trace norm that scales suitably polynomially in $n$, 
then one could find an polynomial time 
algorithm that provides an efficient
classical solution to a ${\tt QMA}$-hard problem,
which leads to a contradiction unless
${\tt QMA}={\tt P}$. Therefore, on the level
of Hamiltonians and accompanying exact
ground states, a full dimension reduction
to one-dimensional problems is in general implausible.}


\subsection{Further numerical results for the ground state block entropy profiles of the half-filled $8\times 8$ spin-less fermion model}

\je{In this section, we elaborate in more detail on the entropy reduction by means of mode transformations as discussed in the main text, to further corroborate our main claims. 
Fig.~5
shows the (von Neumann) entanglement entropy of a half chain for an $N\times N$ lattice for
$N=6$, so $n=N^2$ fermionic modes, as a function of the inverse bond dimension $D$, for several values of the interaction $U$. Depicted are the raw data, as well as a fit 
to the function $x\mapsto y(x)$
defined as
\begin{equation}
y= a + b\frac{1}{x^2} + c\frac{1}{x^4},
\label{eq:fit_even}
\end{equation}
for suitable real $a,b,c$, signifying the interesting regime for large bond dimension $D$. The striking insight is that not only is the entanglement entropy is drastically reduced, compared to the values without mode transformation. But in fact, the values for the entanglement entropy saturate for large bond dimensions $D$, instead of being divergent. This is a convincing illustration of the power of mode transformations to reduce the entanglement entropy in this dimension reduction in description. 
Fig.~6
shows the same plot for $N=8$, with compatible findings.}

\begin{figure}[!htb]
\includegraphics[width=.9\columnwidth]{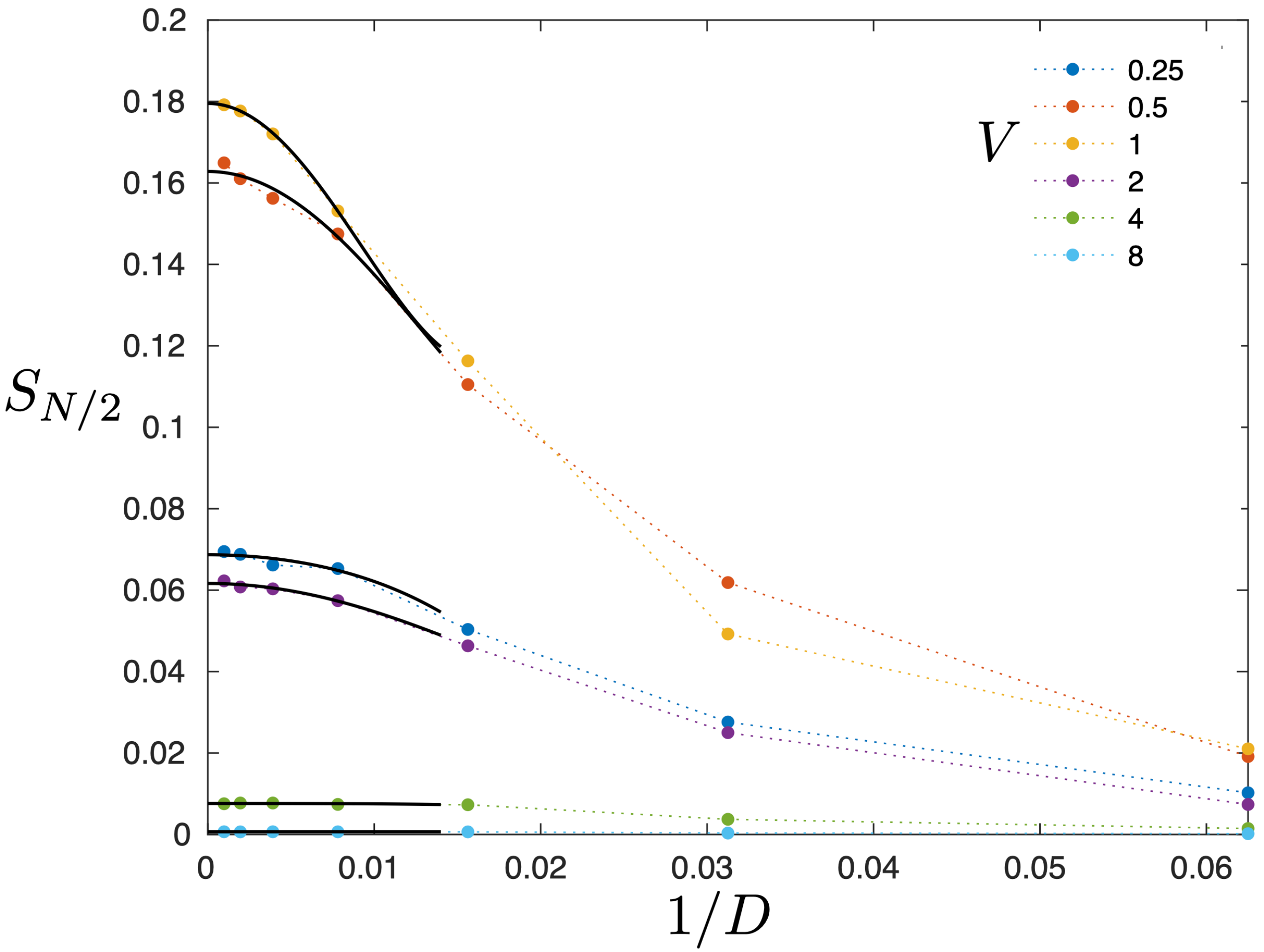}
\caption{\je{Half chain block entropy obtained for the $6\times 6$ spin-less fermion model as a function of inverse bond dimension for various interaction strengths. The solid line is a fit defined by Eq.~(\ref{eq:fit_even})}}.
\label{fig:6x6Entropy}
\end{figure}

\begin{figure}[!htb]
\includegraphics[width=.9\columnwidth]{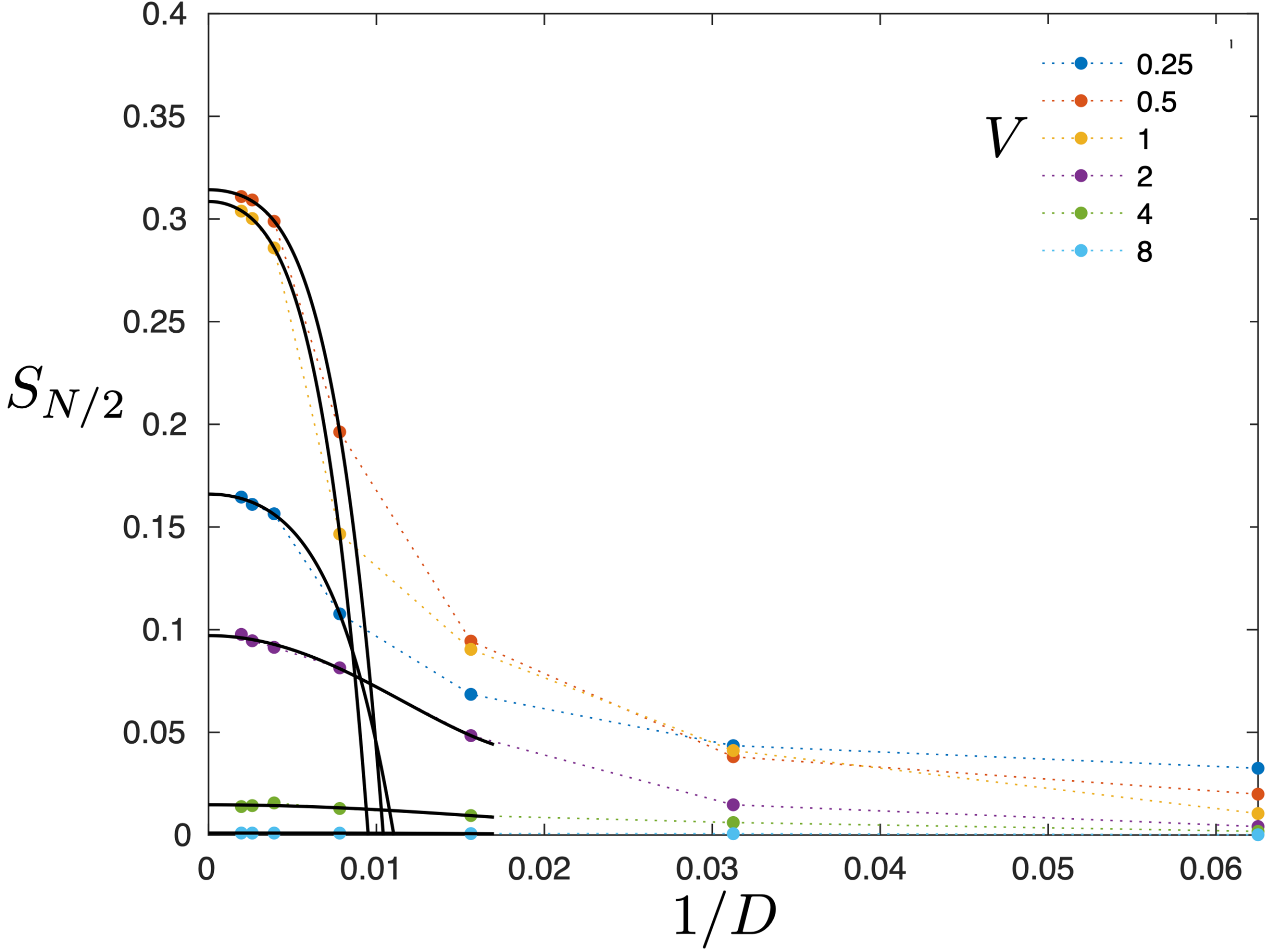}
\caption{\je{Similar to Fig.~5, but for the $8\times 8$ spin-less fermion model.}}
\label{fig:8x8Entropy}
\end{figure}

\subsection{Further computational aspects}

\OL{
The reference real space DMRG data \je{has been} generated by fixing the bond dimension 
\je{$D$} from the very beginning of the DMRG calculations, the residual error threshold in the Lanczos and Davidson diagonalization steps 
\je{has been} set to \je{be} 
$10^{-9}$ and we \je{have} used some $13-15$ sweeps. The maximum value of the truncation error \je{has been} in the range of $10^{-6}$ -- $10^{-7}$. The half-chain \je{(von Neumann)
entanglement entropy} data via mode optimization up to $D=1024$ 
\je{is} shown in Figs.~5 and 6, \je{and it has been} obtained with similar settings, but using $7-9$ sweeps and $60$ iterations \je{of
mode transformations.}

On the practical side, the effective 
\je{Hamiltonian} in the DMRG treatment gets more dense, i.e., additional terms are generated during the curse of
mode optimization which require 
\je{substantial} more computational efforts. \je{However}, the tremendous reduction in the block entropy and the bond dimension 
\je{largely} overcompensate this. In addition, the extra terms that are generated can be applied independently during the diagonalization step.
\je{Thus, the idea of effective dimension reduction by means of mode transformation 
constitutes an ideal candidate for GPU based massive parallelization \cite{Nemes-2014}.} 
}

\subsection{Mode transformation analysis using rotations based on natural orbitals for the half-filled $8\times 8$ spin-less fermion model}

Through the course of basis optimization, the residual 
quantum correlations that have to be captured by the tensor network
ansatz are significantly reduced. As a further proxy for this behaviour,
one may investigate the sum of the single mode von-Neumann entropies
$I_{\rm tot}=\sum_i s_i$ that is reduced drastically, while
pair-wise correlations reflected by $I_{i,j}$ get very much localized
(for additional numerical data see Fig.~\ref{fig:mt-iter-8x8}).
In addition, the investigation of the one-particle reduced density matrix shows that the optimized basis converges to the natural orbital basis 
as $\lambda_i$ and $\langle n_i\rangle$ tend to lie on the top of each 
other (Fig.~\ref{fig:mt-iter-8x8}). Therefore, here the final basis
is the natural orbital basis, but the underlying basis has been systematically
rotated by each mode transformation iterations. 

\begin{figure}[!htb]
\includegraphics[width=1.0\columnwidth]{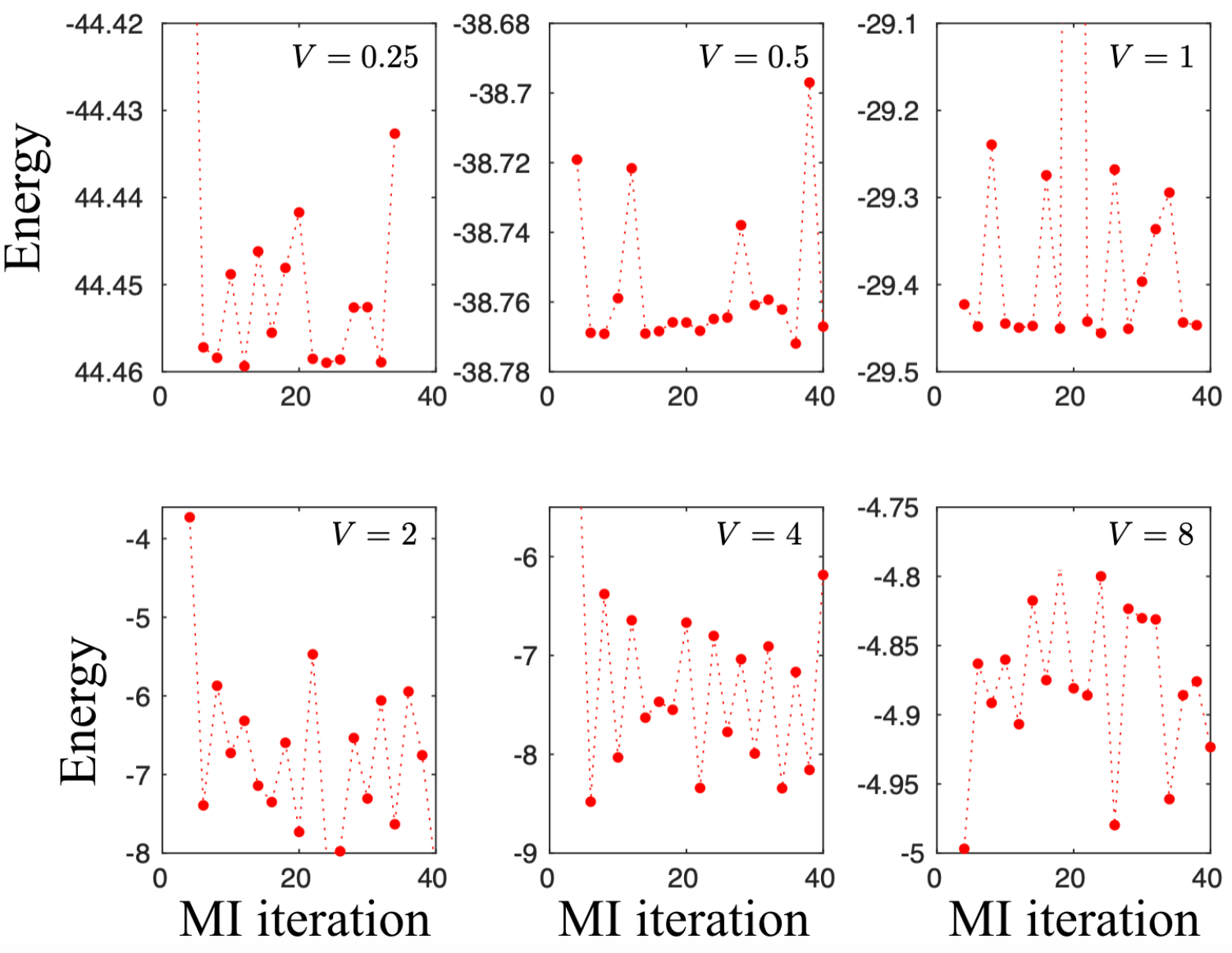}
\caption{Convergence of the ground state energy for the half-filled
$8\times8$ spin-less fermion model as a function of mode transformation iterations 
with fixed bond dimension of 
$D_{\rm opt}=256$ for various $V$ values if we rotate to the natural orbitals after the 7$^{\rm th}$ sweep of each iteration instead of using the local updates and perform another 7 sweeps to obtain a converged ground state in the current rotated basis in order to determine the optimal ordering for the next iteration. Therefore, each iteration based on natural orbitals corresponds to every second iteration based on fermionic mode transformation.}
\label{fig:mt-iter-8x8_NOenergy}
\end{figure}

Since the final basis is the natural orbital basis (see Fig.~\ref{fig:mt-iter-8x8}), 
one might think that a
natural step is to aim at identifying a globally
optimal single particle basis could be more directly based on natural orbitals, i.e., by instead of using the local updates to the single particle basis one could rotate
to the natural orbitals at the end of each mode transformation iteration. Such an
approach has already been tested for quantum chemical 
applications \cite{Rissler-2006}, 
but a very unstable performance has been reported. 
In fact, we have
also found that in the small-$V$ limit such an approach works acceptably,
but for larger $V$ values it breaks down (see Fig.~\ref{fig:mt-iter-8x8_NOenergy}). 
The reason is that for small $V$ the optimal orbitals are Hartree-Fock like 
orbitals, while for large $V$ values localized orbitals seem to be more optimal. Our novel method based on fermionic mode transformation is, however, stable for all $V$ values. Importantly, it can also be used
in general for interacting quantum many body systems.

\newpage
$\,$
\newpage

\begin{widetext}

\subsection{Monitoring various entropic quantities through the course of mode transformations}

\begin{figure}[!htb]
\includegraphics[width=1.0\columnwidth]{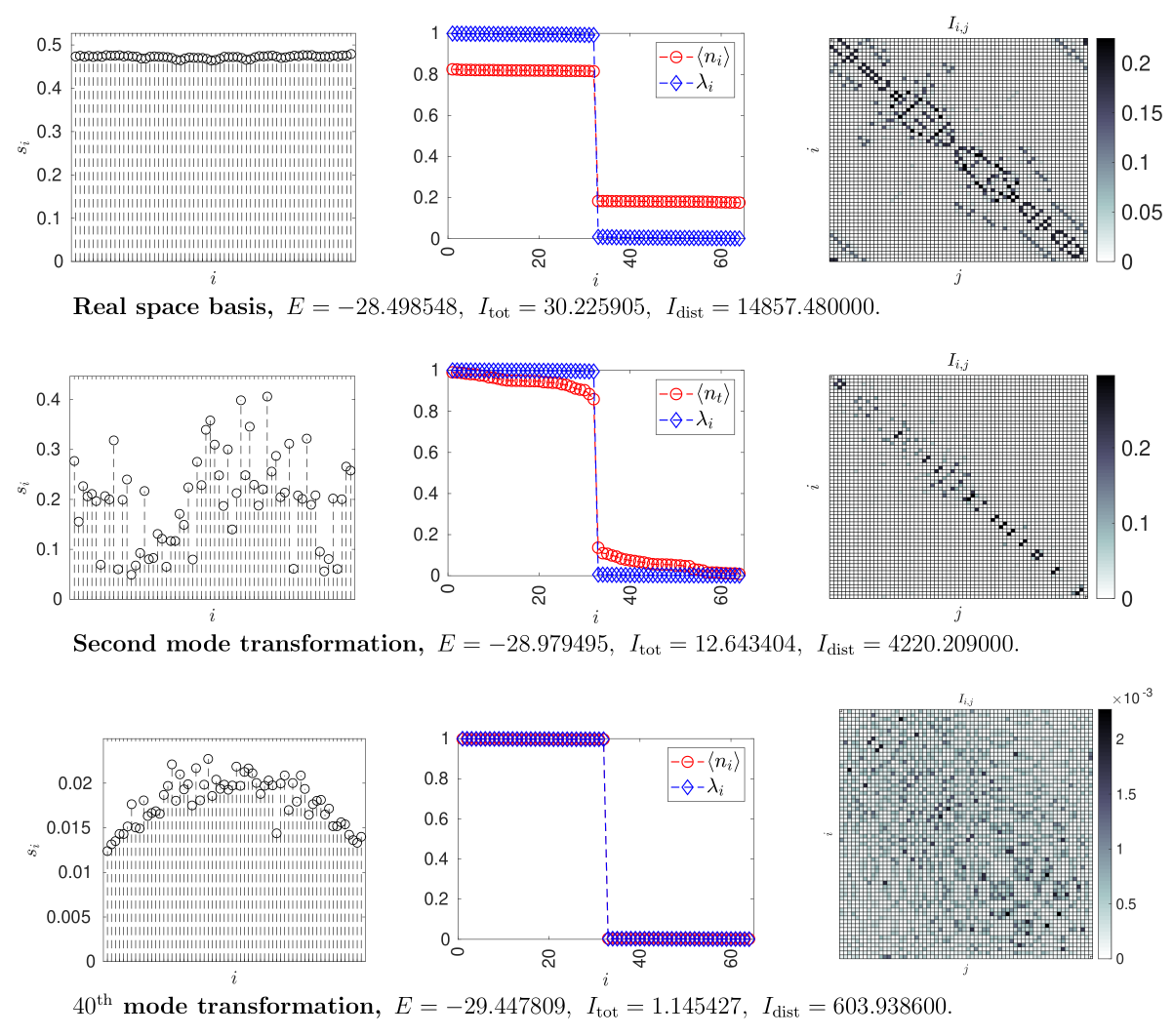}
\caption{Site entropy profiles $\{s_i\}$, 
sorted values of the \je{natural orbital occupation numbers
$\{\lambda_i\}$, occupation numbers $\{\langle n_i\rangle\}$ and mutual 
informations $\{I_{i,j}\}$}
for the real space basis (first row), and
for the 2$^{\rm nd}$ and 40$^{\rm th}$ 
mode transformation iterations  
for the half-filled $8\times8$ spin-less fermion model for $V=1$ and   
$D_{\rm opt}=256$. The ground state energy, the sum of the site 
entropy $I_{\rm tot}$, and the entanglement distance 
$I_{\rm dist}=\sum_{i,j} I_{i,j}|i-j|^2$, 
are printed below the corresponding panels. 
}
\label{fig:mt-iter-8x8}
\end{figure}

\end{widetext}

\end{document}